\documentclass[12pt]{article}
\begin{document}
\title{\textbf{Kantowski-Sachs Universe Models in $f(T)$ Theory of Gravity}}
\author{M. Jamil Amir \thanks{%
mjamil.dgk@gmail.com} and M. Yussouf \thanks{%
yussouf@yahoo.com} \\
Department of Mathematics, University of Sargodha,\\
Sargodha-40100, Pakistan.}
\date{}
\maketitle
\begin{abstract}
The $f(T)$ theory is recently proposed to explain the present cosmic
accelerating expansion of the universe. $f(T)$ theory is an extension of
Teleparallel theory of gravity, where $T$ is the torsion scalar.
This paper contains the construction of $f(T)$ models within the
Kantowski-Sachs universe. For this purpose, we use conservation
equation and equation of state parameter, which represents the
different phases of the universe. We discuss possible cases for
the matter dominated era, radiation dominated era, present dark
energy phase and their combinations. Particularly, a constant
solution has been obtained which may correspond to the
cosmological constant. Further, we consider two well known $f(T)$
models and derive the equation of state parameter and discuss the
cosmic acceleration. Also, the Hubble parameter and average scale
factor have been evaluated.
\end{abstract}

\section{Introduction}

Various cosmological observations, including the type Ia Supernova
[1], the cosmic microwave background radiation [2] and the large
scale structure [3,4], have shown that the universe is undergoing
an accelerating expansion and it entered this accelerating phase
only in the near past. The unexpected observed phenomenon poses
one of the most puzzling problems in cosmology today. Usually, it
is assumed that there exists, in our universe, an exotic energy
component with negative pressure, named dark energy $(DE)$, which
dominates the universe and drives it to an accelerating expansion
at recent times.

Many candidates of DE have been proposed such as the cosmological
constant, quintessence, phantom, quintom as well as the
(generalized) Chaplygin gas, and so on. However, alternatively, we
can take this observed accelerating expansion as a signal of the
breakdown of our understanding to understand the laws of
gravitation. Thus, a modified theory of gravity is needed.
Modified theories of gravity, e.g., Scalar tensor theory,
Brans-Dick theory, String theory, Gauss-Bonnet theory, $f(R)$
theory,  $f(T)$ gravity etc. have recently gained a lot of
interest during the last decade. These theories provide the very natural gravitational
alternative for the DE. The modification of gravitational action
may resolve cosmological problems, paradigm  $DE$ and $DM$ issues. In
this paper we focus our attention only on $f(T)$ theory of
gravity. This theory of gravity is the generalization of
teleparallel theory of gravity [5].

$F(T)$ theory is proposed best to account for the present
accelerating expansion [6-9]. In teleparallel gravity $(TPG)$, we use the
Weitzenb$\ddot{o}$ck connection instead of using the Levi-Civita
connection, which we usually used in $GR$. As a result, in $TPG$, the
Weitzenb$\ddot{o}$ck spacetime has only non-zero torsion and is
curvature free. Similar to $GR$, where the action involves the curvature
scalar $R$, the action of $TPG$ is obtained by simply replacing $R$ with torsion scalar $T$.
In analogy to the $f(R)$ theory, Bengochea and Ferraro suggested
[6] a modified $TPG$ theory, named  $f(T)$  theory, by generalizing the action
of $TPG$,i.e., by replacing $T$ with $f(T)$. They found that it can explain the observed acceleration
of the universe. It is worth mentioning here that the field
equations of $f(R)$ theory are of fourth order while the field equations
of $f(T)$ theory are of second order, which seem easier to be solved.

Linder proposed two new  $f(T)$  models in order to explain the
present cosmic accelerating expansion [7]. He said that  $f(T)$
theory could unify a number of interesting extensions of gravity
beyond $GR$. He investigated that the power law and exponential
models depending upon torsion might give the de-Sitter fate of the
universe. Wu and Yu [10] analyzed the dynamical property of this
theory by using a concrete power law model and showed that the
universe could evolve from radiation dominated era to matter
dominated era and finally enter in an exponential expansion era.

Yang [11] introduced some new $f(T)$  models and gave their
physical implications and cosmological behavior. Wu and Yu [12]
discussed two new  $f(T)$  models and showed how the crossing of
phantom divide line takes place. They also explained the
observation constraints corresponding to these models. Karami and
Abdolmaleki [13] found that equation of state $EoS$ parameter of
holographic and new age graphic  $f(T)$  models always cross the
phantom divide line where entropy connected model has to
experience some conditions on parameters model. The same author
[14] obtained $EoS$ parameter of polytropic, standard, generalized
and modified Chaplygin gas in this modified scenario. Dent, et al.
[15] investigated this theory at the background and perturbed
level and also explored it for quintessence scenarios. Li, et al.
[16] explored local Lorentz invariance and remarked that $f(T)$
theory is not local Lorentz invariant.

Chen, et al. [17] investigated expressions for growth factor,
stability and vector-tensor perturbations. Bamba, et al. [18]
studied the cosmological equations of $EoS$ in exponential,
logarithmic and their combined  $f(T)$ models. Wang [19] searched
spherically symmetric static solution of $f(T)$  models with a
Maxwell term and demonstrated that in conformal Cartesian
coordinates the Reissner-Nordstrom solution does not exist in this
theory. Myrzakulov [20] discussed different $f(T)$ models including
scalar fields and gave analytical solutions for scale factors and
scalar fields.

Sharif and Rani explored Bianchi type-1 universe using
different $f(T)$ gravity models [21]. They also discussed K-essence
models in the framework of  $f(T)$ gravity. Recently,
we explored Kantowski-Sachs universe models in $f(T)$ theory of Gravity [22].
Recently, some interesting  $f(T)$ models have been explored by different authors in [23]-[25].
In this paper, we explore some $f(T)$ models within the
Kantowski-Sachs universe. For this purpose, we use conservation
equation and equation of state parameter, which represent the
different phases of the universe. Also, we discuss the cosmic acceleration of the
universe and $EoS$ parameter by considering two particular $f(T)$ models.

The structure of the paper is as follows. In section $2$, we shall
present some basics of the $f(T)$ theory of gravity and the
corresponding field equations for Kantowski-Sachs spacetime.
Section $3$ contains a detailed construction of $f(T)$ models by
using two different approaches. Section $4$ is devoted to study
the $EoS$ parameter for two particular models and also a discussion
on cosmic acceleration is provided. In the last section, we
summarize and conclude the results.

\section{An Overview of Generalized Teleparallel Theory $f(T)$}

In this section, we introduce briefly the teleparallel theory of
gravity and its generalization to $f(T)$ theory. The Lagrangian
density for teleparallel and $f(T)$ gravity are, respectively,
given as follows [22]:

\begin{eqnarray}
L_T&=&\frac{h}{16 \pi G}T,\\
L_{F(T)}&=&\frac{h}{16 \pi G}F(T),
\end{eqnarray}
where $T$ is the torsion scalar, $f(T)$ is a general
differentiable function of torsion, $G$ is the gravitational
constant and $h=det({h^i}_\mu)$. Mathematically, the torsion
scalar is defined as
\begin{eqnarray}
T= {S_\rho}^{\mu\nu}{T^\rho}_{\mu\nu},
\end{eqnarray}
where  ${S_\rho}^{\mu\nu}$ is antisymmetric in its upper indices
while ${T^\rho}_{\mu\nu}$ is antisymmetric torsion tensor in its
lower  indices. Here ${S_\rho}^{\mu\nu}$ is determined by the
relation
\begin{eqnarray}
S^{\mu\rho\sigma}=\frac{1}{4}(T^{\mu\rho\sigma}+T^{\rho\mu\sigma}-
T^{\sigma\mu\rho})-\frac{1}{2}(g^{\mu\sigma}{T^{\lambda\rho}}_\lambda-g^{\rho\mu}
{T^{\lambda\sigma}}_\lambda)
\end{eqnarray}
and ${T^\lambda}_{\mu\nu}$ is defined as [26]
\begin{eqnarray}
{T^\lambda}_{\mu\nu}={\Gamma^\lambda}_{\nu\mu}-{\Gamma^\lambda}_{\mu\nu}={h^\lambda}_i
\left(\partial_\mu{h^i}_\nu-\partial_\nu{h^i}_\mu\right).
\end{eqnarray}
Here  ${h^i}_\mu$ are the components of the non-trivial tetrad
field $h_i$ in the coordinate basis. It is an arbitrary choice to
choose the tetrad field related to the metric tensor
${g}_{\mu\nu}$ by the following relation
\begin{eqnarray}
g_{\mu\nu}=\eta_{ij} { h^i}_\mu {h^j}_\nu,
\end{eqnarray}
where $\eta_{ij}$ is the Minkowski spacetime for the tangent space
such that $\eta_{ij}=diag(+1,-1,-1,-1)$. For a given metric there
exists infinite different tetrad fields ${h^i}_\mu $ which satisfy
the following properties:
\begin{eqnarray}
{h^i}_\mu{h_j}^\mu={\delta_j}^i;
{h^i}_\mu{h_i}^\nu={\delta_\mu}^\nu.
\end{eqnarray}

In this paper, the Latin alphabets $(i,j,..   =0,1,2,3)$ will be
used to denote the tangent space indices and the Greek alphabets
$(\mu,\nu,...    =0,1,2,3)$ to denote the spacetime indices. The
variation in the indices other than the above mentioned range will
be specified when needed. The variation of Eq.(2) with respect to
the vierbein field leads to the following field equations
\begin{eqnarray}
\left[e^ {-1}
\partial_\mu\left(e{S_i}^{\mu\nu}\right)+{h_i}^\lambda {T^\rho}_{\mu\lambda}
{S_\rho}^{\nu\mu}\right]F_T+{S_i}^{\mu\nu}\partial_\mu(T)F_{TT}\nonumber\\+
\frac{1}{4}{h_i}^\nu F =\frac{1}{2}\kappa^{2}{h_i}^\rho
{T_\rho}^\nu.
\end{eqnarray}
Here $f_T=\frac{df}{dT},  f_{TT}=\frac{d^2 f}{dT^2}, \kappa^2=8\pi
G,{S_i}^{\mu\nu}={h_i}^\rho{S_\rho}^{\mu\nu}$,  and $T_{\mu\nu}$ is
the energy-momentum tensor,given as
\begin{eqnarray}
{T_\rho}^\nu=diag\left(\rho_m, -p_m, -p_m, -p_m\right),
\end{eqnarray}
where $\rho_m$ is the density while $p_m$ is the pressure of the
matter inside the universe.\\
\textbf{ The Field Equations}\\
The line element for a flat, homogeneous and anisotropic
Kantowski-Sachs spacetime is
\begin{eqnarray}
ds^{2}=dt^{2}-A^{2}(t)dr^{2}-B^{2}(t)\left(d\theta^{2}+\sin^{2}\theta d\phi^{2}\right),
\end{eqnarray}
where the scale factors $A$ and $B$ are functions of cosmic time
$t$ only. Using Eqs.$(6)$ and $(10)$, we obtain tetrad components as
follows [27]:
\begin{eqnarray}
{h^i}_\mu&=&diag\left(1,A,B,B\sin\theta\right),\nonumber\\
{h_i}^\mu&=&diag\left(1,A^{-1},B^{-1},(B\sin\theta)^{-1}\right),
\end{eqnarray}
which obviously satisfy Eq.$(7)$. Substituting Eqs.$(4)$ and
$(5)$ in$(3)$ and using$(10)$, it follows after some manipulation
\begin{eqnarray}
T=-2\left(\frac{2\dot{A}\dot{B}}{AB}+\frac{\dot{B}^2}{B^2}\right).
\end{eqnarray}
The field equations $(8)$, for $i=0=\nu$ and $i=1=\nu$, turn out to
be
\begin{eqnarray}
F-4\left(\frac{2\dot{A}\dot{B}}{AB}+\frac{\dot{B}^2}{B^2}\right) F_T=2\kappa^{2}\rho_m,\\
4\left(\frac{\dot{A}\dot{B}}{B}+\frac{A\dot{B}^2}{B^2}+\frac{A\ddot{B}}{B}+\frac{\dot{A}\dot{B}}{AB}\right)F_T
-16\frac{A\dot{B}}{B}\left(\frac{\ddot{A}\dot{B}}{AB}+
\frac{\dot{A}\ddot{B}}{AB}\right.\nonumber\\-\left.\frac{\dot{A}^2\dot{B}}{A^2B}-\frac{\dot{A}\dot{B}^2}{AB^2}
+\frac{\dot{B}\ddot{B}}{B^2}-\frac{\dot{B}^3}{B^3}\right)F_{TT}
-F=2\kappa^2p_m.
\end{eqnarray}
The conservation equation takes the form
\begin{eqnarray}
\dot{\rho_m}
+\left(\frac{\dot{A}}{A}+2\frac{\dot{B}}{B}\right)\left(\rho_m+p_m\right)=0.
\end{eqnarray}
The average scale factor $R$, the mean Hubble parameter $H$ and
the anisotropy parameter $\Delta$ of the expansion respectively
become
\begin{eqnarray}
\dot{\rho_m}
+\left(\frac{\dot{A}}{A}+2\frac{\dot{B}}{B}\right)\left(\rho_m+p_m\right)=0.
\end{eqnarray}
where $H_i$ are the directional parameters in the direction
$x$,$y$ and $z$ respectively given as
\begin{eqnarray} H_1&=&\frac{\dot{A}}{A} , \nonumber\\
H_2&=&\frac{\dot{B}}{B}=H_3.
\end{eqnarray}
It is mentioned here that the isotropic expansion of the universe
is obtained for $\Delta=0$ which further depends
upon the values of unknown scale factors and parameters involved
in the corresponding models [28]-[30].

The equation $(12)$ can be written as
\begin{eqnarray}
2T=J-9H^{2}, J=\frac{\dot{A}}{A}+2\frac{\dot{B}}{B},
\end{eqnarray}
which implies that
\begin{eqnarray}
H=\frac{1}{3}\sqrt{J-2T}.
\end{eqnarray}
If we take $F(T)=T$ then Eqs.$(13)$ and $(14)$ will reduce to
\begin{eqnarray}
\rho_m + \rho T&=&\frac{1}{2\kappa^{2}}\left[-4\left(\frac{2\dot{A}\dot{B}}{AB}+\frac{\dot{B}^2}{B^2}\right)+T\right],
\\
p_m+pT&=&\frac{1}{2\kappa^{2}}\left[4\left(\frac{\dot{A}\dot{B}}{B}+\frac{A\dot{B}^2}{B^2}
+\frac{A\ddot{B}}{B}+\frac{\dot{A}\dot{B}}{AB}\right)-T\right],
\end{eqnarray}
where $\rho T$ and $pT$ are the torsion contributions given respectively as
\begin{eqnarray}
\rho\ T=\frac{1}{2\kappa^{2}}\left[-4\left(\frac{2\dot{A}\dot{B}}{A
B}+\frac{\dot{B}^2}{B^2}\right)\left(1-F_T\right)+T-F\right],
\end{eqnarray}
and
\begin{eqnarray}
pT&=&\frac{1}{2\kappa^{2}}\left[4\left(\frac{\dot{A}\dot{B}}{B}+\frac{A\dot{B}^2}{B^2}+\frac{A\ddot{B}}{B}+\frac{\dot{A}\dot{B}}{AB}\right)
(1- F_T)\right.\nonumber\\&+&16\frac{A\dot{B}}{B}\left(\frac{\ddot{A}\dot{B}}{A B}+
\frac{\dot{A}\ddot{B}}{A
B}-\frac{\dot{A}^2\dot{B}}{A^2B}-\frac{\dot{A}\dot{B}^2}{AB^2}\right.\nonumber\\
&+&\left.\left.\frac{\dot{B}\ddot{B}}{B^2}-\frac{\dot{B}^3}{B^3}\right)F_{TT} -T+F\right].
\end{eqnarray}
The relationship between energy density $\rho$ and pressure of
matter $p$ is described by $EoS$,  $p=\omega \rho$ where $\omega$
is the $EoS$ parameter. For normal, relativistic and
non-relativistic matters, $EoS$ parameter has different
corresponding values. Using Eqs.$(13)$ and $(14)$, the $EoS$
parameter is obtained as follows
\begin{eqnarray}
\omega=-1+\frac{4\left(E-U\right)F_T-16ZF_{TT}}{-4UF_T+F},
\end{eqnarray}
where
\begin{eqnarray}
E&=&\frac{\dot{A}\dot{B}}{B}+\frac{A\dot{B}^2}{B^2}+\frac{A\ddot{B}}{B}+\frac{\dot{A}\dot{B}}{AB},
\\
U&=&\frac{2\dot{A}\dot{B}}{A
B}+\frac{\dot{B}^2}{B^2},
\\
Z&=&\frac{A\dot{B}}{B}\left[\frac{\ddot{A}\dot{B}}{A B}+
\frac{\dot{A}\ddot{B}}{A
B}-\frac{\dot{A}^2\dot{B}}{A^2B}-\frac{\dot{A}\dot{B}^2}{AB^2}
+\frac{\dot{B}\ddot{B}}{B^2}-\frac{\dot{B}^3}{B^3}\right].
\end{eqnarray}
It is mentioned here that the homogeneous part of Eq.$(13)$
yields the following solution
\begin{eqnarray} F(T)=\frac{C_0}{\sqrt{T}},
\end{eqnarray}
where $C_0$ is an integration constant. Using this equation in
Eq.$(14)$, we obtain
\begin{eqnarray}
p_m=-\frac{c_0}{2\kappa^{2}}\left(\frac{2E}{T}+\frac{12Z}{T^2}+1\right)\frac{1}{\sqrt{T}}.
\end{eqnarray}
It is mentioned here that the $p_m$ vanishes for the $FRW$ spacetime [31].\\

\section{Construction of Some $F(T)$ Models}

Here we construct some $F(T)$  models with different cases of
perfect fluid by using two approaches. In the first approach we
use the continuity equation $(15)$ while in the second approach,
$EoS$ parameter $(26)$ will be used. As the constituents of the
universe are non-relativistic matter, radiation and DE, we
consider the
corresponding values of $\omega$ in the following subsections.\\

\subsection{Using Continuity Equation}

In this approach, we use the following relation [32] for
Kantowski- Sachs spacetime
\begin{eqnarray}\frac{1}{9}\left(\frac{\dot{A}}{A}+2\frac{\dot{B}}{B}\right)^2=H_0^2
+\frac{\kappa^{2}\rho_0}{3AB^2\sin\theta},
\end{eqnarray}
where $H_0$ is the Hubble constant having primary implication in
cosmology and $\rho_0$ is an integration constant. The value of $H_0$
corresponds to the rate at which the universe is expanding today.
This equation implies that
\begin{eqnarray}
\left(AB^2\sin\theta\right)^{-1}=\frac{3}{\kappa^{2}\rho_0}\left(H^2-H_0^2\right).
\end{eqnarray}
Using $EoS$ in Eq.$(15)$, it follows that
\begin{eqnarray}\frac{\dot{\rho_m}}{\rho_m}+3H\left(1+\omega\right)=0.
\end{eqnarray}
The components of the universe are described by the terms dark
matter$(DM)$ and dark energy$(DE)$. We consider different cases of
fluids and their combination to construct corresponding $F(T)$
models. For example, for relativistic matter,
$\omega=\frac{1}{3}$, for non relativistic matter, it is zero and
for $DE$ era, it is equal to $-1$ [33].\\
\textbf{ Case 1 $(\omega=0)$:}\\
This is the case of non-relativistic matter, like cold dark
matter ($CDM$) and baryons. It is well approximated as
pressureless dust and called the matter dominated era. Inserting
$\omega=0$ in Eq.$(34)$ and using Eq.$(33)$, we have
\begin{eqnarray}\rho_m=\frac{\rho_c}{AB^2}=\frac{3\rho_c\sin\theta}{\kappa^{2}\rho_0}(H^2-H_0^2),
\end{eqnarray}
where $\rho_c$ is an integration constant. In terms of torsion
scalar, the above equation becomes
\begin{eqnarray}
\rho_m=\frac{\rho_c\sin\theta}{3\kappa^{2}\rho_0}\left(J-9H_0^2-2T\right).
\end{eqnarray}
Substituting the values of $\rho_m$ from Eq.$(36)$ in Eq.$(13)$, we have
\begin{eqnarray}
2TF_T+F=\frac{2\rho_c\sin\theta}{3\kappa^{2}\rho_0}\left(J-9H_0^2-2T\right),
\end{eqnarray}
which has the solution
\begin{eqnarray}
F(T)=\frac{\rho_c\sin\theta}{3\rho_0\sqrt{T}}\int{
\frac{J-9H_0^2-2T}{\sqrt{T}}}dT.
\end{eqnarray}
This will have a unique solution if the value of $J$ is known
which corresponds to the unknown scale factors. Thus for matter
dominated era, we obtain a model in the form of torsion scalar and
Hubble constant.\\
\textbf{Case 2 ($\omega=\frac{1}{3}$):}\\
Here we consider the relativistic matter, like photons and
massless neutrinos with $EoS$ parameter $\omega= \frac{1}{3}$.
This case represents the radiation dominated era of the universe.
Substituting $\omega= \frac{1}{3}$ in Eq.$(13)$ and making use of Eqs.$(20)$ and $(33)$, we obtain
\begin{eqnarray}
\rho_m = \frac{\rho_r\sin^{\frac{4}{3}}\theta}
 {{3}^\frac{4}{3}\kappa^{\frac{8}{3}}\rho^{\frac{4}{3}}_0}\left(J-9H_0^2-2T\right)^{\frac{4}{3}},
\end{eqnarray}
where $\rho_r$ is another integration constant. Inserting this value of
$\rho_m$ in Eq.$(13)$, we get
\begin{eqnarray}2TF_T+F= \frac{2\rho
_r\sin^{\frac{4}{3}}\theta}
 {{3}^\frac{4}{3}\kappa^{\frac{2}{3}}\rho^{\frac{4}{3}}_0}\left(J-9H_0^2-2T\right)^{\frac{4}{3}},
\end{eqnarray}
which has solution
\begin{eqnarray}F(T)= \frac{\rho
_r\sin^{\frac{4}{3}}\theta}
 {{3}^\frac{4}{3}\kappa^{\frac{2}{3}}\rho^{\frac{4}{3}}_0\sqrt{T}}
 \int\frac{\left(J-9H_0^2-2T\right)^{\frac{4}{3}}}{\sqrt{T}}dT.
\end{eqnarray}
This also depends upon the value of $J$ as well as torsion scalar
and Hubble constant.\\
 \textbf{Case 3 ($\omega$=-1):}\\
This case represents the present $DE$ constituting $74$ percent of
the universal density. $DE$ is assumed to have a large negative
pressure in order to explain the observed acceleration of the
universe. It is also termed as energy density of vacuum or
cosmological constant $\Lambda$. Replacing $\omega$=-1 in
Eq. $(34)$, we get
\begin{eqnarray}
\rho_m=\rho_d,
\end{eqnarray}
where $\rho_d$ is an integration constant. Consequently, Eq.$(13)$ takes the form
\begin{eqnarray}
2TF_T+F=2\kappa^{2}\rho_d
\end{eqnarray}
with solution
\begin{eqnarray}
F(T)=\frac{\kappa^{2}\rho_d}{\sqrt{T}}\int\frac{1}{\sqrt{T}}dT.
\end{eqnarray}\\
\textbf{Case 4  (Combination of $\omega=0$ and $\omega=\frac{1}{3}$):}\\
Let us now consider the case when the energy density is a combination of different
fluids, the dust fluid and the radiations. Adding Eqs.$(36)$ and $(39)$,
after simplification, it follows that
\begin{eqnarray}
\rho_m=\frac{\rho_c
\sin\theta}{6\kappa^{2}\rho_0}\left(J-9H_0^2-2T\right)+\frac{\rho
_r\sin^{\frac{4}{3}}\theta}
 {{2}.{3}^\frac{4}{3}\kappa^{\frac{8}{3}}\rho^\frac{4}{3}_0}\left(J-9H_0^2-2T\right)^{\frac{4}{3}}.
\end{eqnarray}
Substituting this value of $\rho_m$ in Eq.$(13)$,  we get
\begin{eqnarray}
2TF_T+F=\frac{\rho_c\sin\theta}{3\rho_0}\left(J-9H_0^2-2T\right)+\frac{\rho
_r\sin^{\frac{4}{3}}\theta}
 {{3}^\frac{4}{3}\kappa^{\frac{2}{3}}\rho^\frac{4}{3}_0}\left(J-9H_0^2-2T\right)^{\frac{4}{3}}
\end{eqnarray}
and its solution is
\begin{eqnarray}
F(T)&=&\frac{\rho_c\sin\theta}
{6\rho_0\sqrt{T}}\int\frac{\left(J-9H_0^2-2T\right)}{\sqrt{T}}dT\nonumber\\&+&\frac{\rho
_r\sin^{\frac{4}{3}}\theta}
 {{2}.{3}^\frac{4}{3}\kappa^{\frac{8}{3}}\rho^{\frac{4}{3}}_0\sqrt{T}}
 \int \frac{\left(J-9H_0^2-2T\right)^{\frac{4}{3}}}{\sqrt{T}}dT.
\end{eqnarray}\\

\textbf{Case 5 ( Combination of $ \omega =0 $ and $ \omega =-1
$):}\\
The combination of $EoS$ parameters for matter dominated era
and $DE$ yields
\begin{eqnarray}
\rho_m=\frac{\rho_c\sin\theta}{6\kappa^{2}\rho_0}\left(J-9H_0^2-2T\right)+\frac{\rho_d}{2}.
\end{eqnarray}
Inserting this value of $\rho_m$ in Eq.$(13)$, we get
\begin{eqnarray}
2TF_T+F=\frac{\rho_c\sin\theta}{3\rho_0}\left(J-9H_0^2-2T\right)+\kappa^{2}\rho_d,
\end{eqnarray}
yielding
\begin{eqnarray}
F(T)=\frac{\rho_c\sin\theta}
{6\rho_0\sqrt{T}}\int\frac{\left(J-9H_0^2-2T\right)}{\sqrt{T}}dT+\frac{\kappa^{2}\rho_d}{2\sqrt{T}}\int\frac{1}{\sqrt{T}}dT.
\end{eqnarray}\\
\textbf{ Case 6 (Combination of $\omega=-1$ and
$\omega=\frac{1}{3}$):}\\
This case gives the following form of the energy density
\begin{eqnarray}
\rho_m=\frac{\rho_r \sin{^\frac{4}{3}}\theta} {{2}.{3}^\frac{4}{3}
{2}\kappa^{\frac{8}{3}}\rho^{\frac{4}{3}}_0}\left(J-9H_0^2-2T\right)^{\frac{4}{3}}+\frac{\rho_d}{2}.
\end{eqnarray}
Substituting this value in Eq.$(13)$, we get
\begin{eqnarray}
2TF_T+F =\frac{\rho_r\sin^{\frac{4}{3}}\theta}
 {{3}^\frac{4}{3}\kappa^{\frac{4}{3}}\rho^{\frac{4}{3}}_0}\left(J-9H_0^2-2T\right)^{\frac{4}{3}}+\kappa^2 \rho_d,
\end{eqnarray}
which gives
\begin{eqnarray}
F(T)=\frac{\rho_r\sin^{\frac{4}{3}}\theta}
 {{2}.{3}^
 {\frac{4}{3}} \kappa^{\frac{2}{3}}\rho^{\frac{4}{3}}_0\sqrt{T}}
 \int\frac{\left(J-9H_0^2-2T\right)^{\frac{4}{3}}
}{\sqrt{T}}dT+\frac{\kappa^{2}\rho_d}{2\sqrt{T}}\int\frac{1}{\sqrt{T}}dT.
\end{eqnarray}
It is mentioned here that the cases $4$-$6$ provide $F(T)$ models
for combination of different matters. Normally, the dark matter
and DE developed independently. However, there are attempts
[34] to include an interaction amongst them so that one can
get some insights and see the combined effect of different fluids.
Dark matter plays a central role in galaxy evolution and has
measurable effects on the anisotropies observed in the cosmic
microwave background. Although, matter made a large fraction of
total energy of the universe but its contribution would fall in
the far future as DE becomes more dominated. It may provide
an interaction between dark matter and DE and drive
transition from an early matter dominated era to a phase of
accelerated expansion. Using the same phenomenon, DE and
different forms of matter are discussed in the framework of $F(T)$
theory which may help to discuss accelerated expansion of the
universe.\\

\subsection{Using $EoS$ Parameter}

Here we formulate some  $F(T)$ models in a slightly different way.
We substitute different values of parameter $\omega$ in Eq.$(13)$ and solve
it accordingly. The Eq.$(26)$ can be written as
\begin{eqnarray}
16ZF_{TT}-4\left(E+\omega U\right)F_T+\left(1+\omega\right)F=0.
\end{eqnarray}
Now, we construct  $F(T)$ models in the following cases
:\\
\textbf{ Case 1:}\\
When we put $\omega=\frac{1}{3}$ in Eq.$(54)$, we obtain
\begin{eqnarray}
Z F_{TT}-\left(E+\frac{U}{3}\right)F_T+\frac{1}{3}F=0.
\end{eqnarray}
This has the following general solution.
\begin{eqnarray}
F(T)&=& C_1\texttt{ exp} \left[\left\{\frac{\left(3E+U\right)+\sqrt{\left(3E+U\right)^2-12Z}}{6Z}\right\}T\right]
\nonumber\\ & +&C_2\texttt{ exp} \left[\left\{\frac{\left(3E+U\right)-\sqrt{(3E+U)^2-12Z}}{6Z}\right\}T\right],
\end{eqnarray}
where $C_1$ and $C_2$ are constants. \\
\textbf{Case 2:} \\
Here we consider the dust case when pressure is zero, that is, $\omega$ =0. Then the Eq.$(52)$ takes the form
\begin{eqnarray}
16ZF_{TT}-4EF_T+F=0.
\end{eqnarray}
It has the following general solution
\begin{eqnarray}
F(T)&=&C_3\texttt{ exp} \left[\left\{\frac{E+\sqrt{E^2-4Z}}{8Z}\right\}T\right]\nonumber\\&+&
 C_4\texttt{ exp }\left[\left\{\frac{E-\sqrt{E^2-4Z}}{8Z}\right\}T\right],
\end{eqnarray}\\
where $C_3$ and $C_4$ are constants.\\
\textbf{ Case 3:}\\
For $\omega=-1$,  Eq.$(54)$ becomes
\begin{eqnarray}
4ZF_{TT}-\left(E-U\right)F_T=0,
\end{eqnarray}
whose general solution is
\begin{eqnarray}
F(T)=C_5+C_6\texttt{ exp} \left[\left(\frac{E-U}{4Z}\right)T\right],
\end{eqnarray}
where $C_5$ and $C_6$ are constants. The Eqs.$(54)$, $(56)$
and $(58)$ represent $F(T)$ models corresponding  radiation,
matter and DE phases respectively. The exponential form of $F(T)$
models represents a universe which always lies in phantom or
non-phantom phase depending on parameters of the models [35].\\

\section{Construction of $EoS$ Parameters and \\Cosmic Acceleration }

In this section we derive $EoS$ parameter by using two different
$F(T)$ models and also investigate cosmic acceleration. For this
purpose, we evaluate $\rho_{m}$ and $p_{m}$ using the field
equations and then construct the corresponding $EoS$ parameters. \\

\subsection{The First Model}

Consider the following $F(T)$ model [31]
\begin{eqnarray}
F= \alpha T + \frac{\beta}{T},
\end{eqnarray}
where $\alpha$ and $\beta$ are positive real constants. Inserting this
value of $F$ in Eqs.$(13)$ and $(14)$, it follows that\\
\begin{eqnarray}
2 \kappa ^{2} \rho_{m}=\left(-4U+T\right)\alpha + \beta \left(1+4UT^{-1}\right)T^{-1}, \\
2 \kappa ^{2} p_{m}= \left(4E-T\right)\alpha - \beta
\left(4ET^{-1}+32ZT^{-2}+1\right)T^{-1}.
\end{eqnarray}
Dividing Eq.$(61)$ by $(60)$, the $EoS$ parameter is obtained as
follows
\begin{eqnarray}
\omega=-1+\frac{4\left(E-U\right)\alpha - \beta
\left(4\left(E-U\right)T^{-1}+32ZT^{-1}\right)T^{-2}}{\left(-4U+T\right)\alpha + \beta
\left(1+4UT^{-1}\right)T^{-1}}.
\end{eqnarray}
Now, we would like to discuss the last equation for particular values of $\alpha$ and
$\beta$. For $\alpha$ $\neq$0, $\beta=0$, we obtain 
\begin{eqnarray}
\omega=-1+\frac{2}{3}\left(1-\frac{E}{U}\right).
\end{eqnarray}
This leads to three different cases of $\omega$ representing
different phases of the evolution of the universe as follows:
\begin{itemize}
\item  If $\frac{E}{U} >1$ then  $\omega <-1$,
which corresponds to the phantom accelerating universe.
\item  When $\frac{E}{U} <1$ then  $\omega >-1$, slightly
which corresponds to the quintessence region.
\item  When $\frac{E}{U}=1$, we obtain a universe whose dynamics is
 dominated by cosmological constant with  $\omega =-1$
which corresponds to the phantom accelerating universe.
\end{itemize}
It is interesting to mention here that model $(59)$ reduces to $GR$
spatially flat Friedmann equation in the limiting case when
anisotropy vanishes. Also, for the case, when $\alpha$ $\neq$0, $\beta$ $\neq$0, we obtain no physical results.

\subsection{The Second Model}

Assume the $F(T)$ has the form [31]
\begin{eqnarray}
F= \alpha T + \beta T^{n},
\end{eqnarray}
where $n$ is a positive real number. The corresponding field
equations become.
\begin{eqnarray}
2 \kappa ^{2} \rho_{m}&=& \left(-4U+T\right)\alpha + \beta \left(-4nUT^{-1}+1\right)T^{n},
\\
2 \kappa ^{2} p_{m}&=& \left(4E-T\right)\alpha +4n \beta E
T^{n-1}-16n(n-1)\beta Z T^{n-2}-\beta T^{n}.
\end{eqnarray}
Consequently, the $EoS$ parameter takes the form
\begin{eqnarray}
\omega=-1+\frac{4\left(-U+E\right)\alpha +4n \beta \left(-U+E\right)T^{n-1}-16n(n-1)\beta
Z T^{n-2}}{\left(-4U+T\right)\alpha + \beta \left(-4nUT^{-1}+1\right)T^{n}}.
\end{eqnarray}
The case $\alpha$ $\neq0$, $\beta=0$, leads to the same discussion
as in the first case. For $\alpha=0$, $\beta \neq 0$, we have
\begin{eqnarray}
\omega=-1+\frac{2n}{2n+1}\left[1- \left\{\frac{E}{U}+ \frac{8n(n-1)Z}{U^{2}}\right\}\right].
\end{eqnarray}
For any positive real number $n$, we can discuss as follows:
\begin{itemize}
\item  When  $\left[\frac{E}{U}+ \frac{8n(n-1)Z}{U^{2}}\right] <1$,
the Eq.$(70)$ gives $\omega <-1$ which represents the phantom
accelerating universe.
\item  For   $\left[\frac{E}{U}+\frac{8n(n-1)Z}{U^{2}}\right]=1$, we obtain
$\omega=-1$ and hence the universe rests in DE era dominated by cosmological
constant.
\item  The case  $\left[ \frac{E}{U}+ \frac{8n(n-1)Z}{U^{2}}\right]<-1$,
corresponds to the quintessence era because $\omega>-1.$
\end{itemize}
Assuming $n=1$ as a particular case in Eqs.$(67)$ and $(68)$, we
have
\begin{eqnarray}
\rho_m&=&\frac{\left(\alpha+\beta\right)\left(-4U+T\right)}{2\kappa^{2}},
\\
p_m&=&\frac{\left(\alpha+\beta\right)\left(4E-T\right)}{2\kappa^{2}}.
\end{eqnarray}
In the following, we discuss the evolution of the scale factor for
Kantowski-Sachs universe. For this purpose, we assume [31]
\begin{eqnarray}
p_m=\frac{A_{-1}(T)}{\rho_m}+A_0(T)+A_1(T)\rho_m,
\end{eqnarray}
such that $A_{-1}$, $A_0$,  $A_1$ are constants. Substituting
Eqs.$(71)$ and $(72)$ in the above equation, it follows that
\begin{eqnarray}
4E-T=\frac{a}{-4U+T}+b+c\left(-4U+T\right),
\end{eqnarray}
where
\begin{eqnarray}
a=\frac{4\kappa^{4}A_{-1}}{\left(\alpha+\beta\right)^{2}},~~~
b=\frac{2\kappa^{2}A_0}{\alpha+\beta},~~~c=A_1.
\end{eqnarray}
This leads to
\begin{eqnarray}
T&=&-\frac{4U+4E-b+8Uc}{2\left(1+c\right)}\nonumber\\
&\pm& \frac{1}{2\left(1+c\right)}\left[\left(4U+4E-b+8Uc\right)^{2}\right.\nonumber\\&-&\left.4\left(1+c\right)\left(16cU^{2}-4bU+a+16UE\right)\right]^{
\frac{1}{2}}.\nonumber\\
\end{eqnarray}
Substituting this value of torsion in Eq.$(21)$ we have
\begin{eqnarray}
H&=&\frac{1}{3}\left[\left|J-\frac{4U+4E-b+8Uc}{1+c}\right.\right.\nonumber\\
&\pm&
\frac{1}{1+c}\left\{\left(4E+4U-b+8Uc\right)^2\right.\nonumber\\&-&\left.\left.\left.4\left(1+c\right)\left(16cU^2-4bU+a+16UE\right)\right\}^
{\frac{1}{2}}\right|\right]^{\frac{1}{2}}.\nonumber\\
\end{eqnarray}
The corresponding average scale factor becomes
\begin{eqnarray}
R&=&R_0 \texttt{exp} \left\{\frac{1}{3}\int \left[\left|J-\frac{4U+4E-b+8Uc}{1+c}\right.\right.\right.\nonumber\\&\pm&
\frac{1}{1+c}\left\{\left(4E+4U-b+8Uc\right)^2\right.\nonumber\\&-&\left.\left.\left.\left.
4\left(1+c\right)\left(16cU^2-4bU+a+16UE\right)\right\}^{\frac{1}{2}}\right|\right]^{\frac{1}{2}}dT\right\}.
\end{eqnarray}
As a special case of model $(73)$, if we take $A_{-1}$ as a
constant while $A_0=0=A_1$, we obtain standard Chaplygin gas $EoS$
[36]. In this respect, Eqs.$(74)$ and $(75)$ give the following
results respectively.
\begin{eqnarray}
T&=&2\left(E+U\right)\pm \sqrt{\left\{2\left(E+U\right)\right\}^2-\left(a+16UE\right)},
\\
H&=&\frac{1}{3}\left[J-4\left(E+U\right)\pm
2\sqrt{\left\{2\left(E+U\right)\right\}^2-\left(a+16UE\right)}\right].
\end{eqnarray}
The average scale factor for Chaplygin gas has the form
\begin{eqnarray}
R=R_0 \texttt{exp} \left\{\frac{1}{3}\int \left| J-4\left(E+U\right)\pm 2\sqrt{\{2\left(E+U\right)\}^2-\left(a+16UE\right)}\right|
dT \right\}.
\end{eqnarray}
This represents an exponential expansion which may result a rapid
increment between the distance of the two non-accelerating
observers as compared to the speed of light. As a result, both
observers are unable to contact each other. Thus if our universe
is forthcoming to a de-Sitter Universe [7], then we would not be
able to observe any galaxy other than our own Milky way system.

\section{Summary and Conclusion}

The study of cosmological models has become burning issue since
the last decade. Much interest has been given by the researchers
to resolve the cosmological problems including the existence of $DE$
and $DM$ in the universe. As $GR$ can not explain the rushing growth
of the universe so we need some other framework of gravity, which
may resolve this issue. There are many alternate theories of
gravity among which $F(T)$ theory of gravity is one of the
candidates.

The purpose of this paper is to investigate the recently developed
F(T) gravity. For this purpose we have taken Kantowski-Sachs
spacetime model describing anisotropic and spherically homogeneous
universe. Some $F(T)$ models have been constructed by using two
different approaches. In the first approach, we have used the
continuity equation while in the second method, $EoS$ is used.  The
results obtained so far in these approaches are given in the
following tables $(1-2)$:

\vspace{0.5cm}

{\bf {\small Table 1.} {\small Expressions for $F(T)$ using Continuity Equation }}
\begin{center}
\begin{tabular}{|c|c|}
\hline{\bf CASES}&{\bf  $F(T)$ }\\
\hline{$1$} &

$\frac{\rho_c\sin\theta}{3\rho_0\sqrt{T}}\int{
\frac{J-9H_0^2-2T}{\sqrt{T}}}dT$\\

\hline{ $2$} &

$\frac{\rho
_r\sin^{\frac{4}{3}}\theta}
 {{3}^\frac{4}{3}\kappa^{\frac{2}{3}}\rho^{\frac{4}{3}}_0\sqrt{T}}
 \int\frac{(J-9H_0^2-2T)^{\frac{4}{3}}}{\sqrt{T}}dT$\\
\hline{ $3$} &

$\frac{\kappa^{2}\rho_d}{\sqrt{T}}\int\frac{1}{\sqrt{T}}dT $\\ \hline{ $4$} &

$\frac{\rho_c\sin\theta}
{6\rho_0\sqrt{T}}\int\frac{(J-9H_0^2-2T)}{\sqrt{T}}dT+\frac{\rho
_r\sin^{\frac{4}{3}}\theta}
 {{2}.{3}^\frac{4}{3}\kappa^{\frac{8}{3}}\rho^{\frac{4}{3}}_0\sqrt{T}}
 \int \frac{(J-9H_0^2-2T)^{\frac{4}{3}}}{\sqrt{T}}dT $\\ \hline{ $5$} &

$\frac{\rho_c\sin\theta}
{6\rho_0\sqrt{T}}\int\frac{(J-9H_0^2-2T)}{\sqrt{T}}dT+\frac{\kappa^{2}\rho_d}{2\sqrt{T}}\int\frac{1}{\sqrt{T}}dT $\\
\hline{ $6$} &

$\frac{\rho_r\sin^{\frac{4}{3}}\theta}
 {{2}.{3}^
 {\frac{4}{3}} \kappa^{\frac{2}{3}}\rho^{\frac{4}{3}}_0\sqrt{T}}
 \int\frac{(J-9H_0^2-2T)^{\frac{4}{3}}
}{\sqrt{T}}dT+\frac{\kappa^{2}\rho_d}{2\sqrt{T}}\int\frac{1}{\sqrt{T}}dT $\\ \hline

\end{tabular}
\end{center}

\vspace{0.5cm}

{\bf {\small Table 2.} {\small Expressions for $F(T)$ using $EoS$ Parameter }}
\begin{center}
\begin{tabular}{|c|c|}
\hline{\bf CASES}&{\bf  $F(T)$}\\
\hline $1$ & $ C_1\texttt{ exp}
[\{\frac{(3E+U)+\sqrt{(3E+U)^2-12Z}}{6Z}\}T]$\\&
 $+C_2\texttt{ exp} [\{\frac{(3E+U)-\sqrt{(3E+U)^2-12Z}}{6Z}\}T]
$\\

\hline{ $2$} &
$C_3\texttt{ exp} [\{\frac{E+\sqrt{E^2-4Z}}{8Z}\}T]+
 C_4\texttt{ exp }[\{\frac{E-\sqrt{E^2-4Z}}{8Z}\}T] $\\ \hline
 \hline{ $3$} &
$C_5+C_6\texttt{ exp} [(\frac{E-U}{4Z})T] $\\ \hline
\end{tabular}
\end{center}
These $F(T)$ gravity models represent three different eras of the
universe corresponding to different values of $EoS$ parameter. These
are the matter, radiation and DE dominated eras corresponding to
$\omega=0$, $\omega=\frac{1}{3}$ and $\omega=-1$ respectively,
given in table $1$ as cases $1$-$3$. If we consider combination of
radiation and matter, we may have more interesting results to
study the developing universe. Using different combinations of $EoS$
parameter we obtain three more models, given in table $1$ as cases
$4$-$6$. Also we have obtained $F(T)$ models in exponential form for
some particular values of $EoS$ parameter, given in table $2$.

It is well known that the evolution of $EoS$ parameter is one of the
biggest efforts in the observational cosmology today. We have
considered two well known $F(T)$ models, given in Eqs.$(61)$ and $(66)$
and found the corresponding expressions for $EoS$ parameter
$\omega$. These expressions have been investigated for some
particular values of the parameters $\alpha$ and $\beta$ which
yield fruitful results corresponding to realistic situations.
Further, we discuss the cosmic acceleration for these models. We
conclude that our universe would approach to de-Sitter universe in
the infinite future. The isotropic expansion of the universe is
obtained for $\Delta=0$ which depends upon the values of unknown
scale factors and parameters involved in the corresponding models.

\vspace{1.5cm}

{\bf References}

\begin{description}

\item{[1]} Riess, A.G.  et al.: Astron. J. \textbf{116}(1998)1009, Perlmutter, S.  et al.:
 Astrophys. J. \textbf{517}(1999)565.
\item{[2]}   Spergel,  D.E. et al.: ApJS. \textbf{175}(2003)148.

\item{[3]} Tegmark, M.: Phys. Rev. \textbf{D69}(2004)103501.

\item{[4]} Einstein, D.J. et al.,:  Astrophys. J.\textbf{633}(2005)560.
\item{[5]} Sharif, M. and Amir,  M.J.: Mod. Phys. Lett.  \textbf{A22}(2007)425; Sharif, M.
and Amir,  M.J.: Gen. Relativ Gravit.
\textbf{39}(2007)989; Sharif, M and Amir M.J.: Int. J. Theor.
Phys. \textbf{47}(2008)1742; Sharif, M. and Amir M.J.: Mod. Phys.
Lett. \textbf{A37}(2007)1292; Sharif, M. and Nazir K.: Commun
Theor. Phys. \textbf{50}(2008)664; Sharif, M. and Taj, S.:
Astrophys. Space Sci.  \textbf{75}(2010)325; Hayashi, K. and
Shirafuji, T.: Phys. Rev.  \textbf{D19}(1979)3524; Sharif, M. and
Amir,  M.J.: Mod. Phys. Lett.  \textbf{A23}(2008)963.
\item{[6]}  Bengochea, G.R and Ferraro, R.: Phys. Rev. \textbf{D79}(2009)124019.

\item{[7]} Linder, E.V.: Phys. Rev. \textbf{D81}(2010)127301.

\item{[8]} R. Myrzakulov, arXiv : 1006.1120. , K.K. Yerzhanov,
S.R. Myrzakulo, I.I. Kulnazarov, R. Kulnazarov, arXiv : 1006.389.,
Wu, P.  and  Yu, H. : Phys. Lett. \textbf{B692}(2010)176., R.
Yang, arXiv : 1007.3571, P.  Yu. Isyba, I.I. Kulnazarov,  K.K.
Yerzhanov,  R. Myrzakulov. arXiv : 1008.0779. , J.B. Dent, S.
Dutta, E.N. Saridakis, arXiv :1008.3188. , P. Wu and H. Yu,
arXiv :1008.3669.

\item{[9]} Bamba, K. , Geng,  C.Q. and Lee, C.C.: arXiv: [astro-ph.] 1008.4036.

\item{[10]} Wu, P. and Yu, H.: Phys. Lett. \textbf{B692}(2010)176.

\item{[11]} Yang, R.J.: Eur. Phys. J. \textbf{C71}(2011)1797.

\item{[12]} Wu, U. and Yu, H. : Eur. Phys. J. \textbf{C71}(2011)1552.
\item{[13]} Karami, K. and Abdolmaleki, A.: Research in Astron. Astrophys. \textbf{13}(2013)757.

\item{[14]}  Karami, K. and Abdolmaleki, A.: Journal of Physics: Conference Series \textbf{375}(2012)032009.

\item{[15]} Dent, J.B. , Dutta, S. and  Saridakis, E.N.: JCAP \textbf{1101}(2011)009.

\item{[16]} Li, B. , Sotiriou, T.P. and Barrow, J.D.: Phys. Rev. \textbf{D83}(2011)064035.

\item{[17]} Chen, S.H. et al.: Phys. Rev. \textbf{D83}(2011)023508.

\item{[18]} Bamba, K. et al.: JCAP \textbf{1101}(2011)021.

\item{[19]} Wang, T.: Phys. Rev. \textbf{D84}(2011)024042.

\item{[20]} Myrzakulov, R.:\textit{ According to cosmology in
            F(T)  Gravity with scalar field},  arXiv/1006.3879.

\item{[21]} Sharif, M. and Rani,  S.: Mod. Phys. Lett. \textbf{A26}(2011)1657.
\item{[22]} Li, B. et al.: Phys. Rev. \textbf{D83}(2011)064035.

 \item{[23]} Cardone, V.F., Radicella, N. and Camera, S.: Phys. Rev. \textbf{D} (to appear).

\item{[24]} Nashed, G.G.L.: arXiv/1403.6937v1.

\item{[25]} Aghamohammadi, A.: arxiv/1402.2607v1.

\item{[26]} Aldrovandi, R. and Pereira, J.G.: \textit{An introduction to
            Geometrical Physics} (World scientific,1995).

\item{[27]} Vakili, B. and Sepangi, H.R.: JCAP \textbf{09}(2005)008.

\item{[28]} Sharif, M. and Zubair, M.: Astrophys. Space Sci. \textbf{330}(2010)399.

\item{[29]} Sharif, M. and Kausar, H.R.:Phys. Lett. \textbf{B697}(2011)1.

\item{[30]} Tiwari, R.K.:\textit{ Research in Astron Astrophys.} \textbf{10}(2010)291.

\item{[31]} Myrzakulov, R.: Eur. Phys. J. \textbf{C71}(2011)1752.

\item{[32]} Elizalde, E. et al.:\textit{ Class. Quantum Grav}. \textbf{27}(2010)095007.

\item{[33]} Bean, R.:\textit{ Mod. TASI Lectures on Cosmic Acceleration}, arXiv/1003.4468.

\item{[34]} Chimento, L.P., Jakubi, A.S. and Zimdahl, W.: Phys. Rev. \textbf{D67}(2003)083513.

\item{[35]} Bamba, K. et al.: JCAP \textbf{01}(2011)021.

\item{[36]} Bilic, N. et al.: J. Phys. \textbf{A40}( 2007)6877.
\end{description}
\end{document}